\documentstyle[aps]{revtex}
\begin{document}
\draft
\title{Transport Properties, Thermodynamic Properties, and Electronic
Structure of SrRuO$_3$}
 
\author{
P. B. Allen$^{(a),(c)}$\cite{address},
H. Berger$^{(b)}$,
O. Chauvet$^{(b)}$,
L. Forro$^{(b)}$,
T. Jarlborg$^{(c)}$,
A. Junod$^{(c)}$,
B. Revaz$^{(c)}$, and
G. Santi$^{(c)}$}
\address{
(a) Institut Romand de Recherche Num\'erique en Physique des Materiaux (IRRMA),\\
IN-Ecublens, CH-1015 Lausanne, Switzerland\\
(b) \'Ecole Polytechnique Federale de Lausanne, Departement de Physique,\\
CH-1015 Lausanne, Switzerland\\
(c) Departement de Physique de la Mati\`ere Condens\'ee, Universit\'e de
Gen\`eve,\\
24 Quai Ernest-Ansermet, CH-1211 Geneva 4, Switzerland}

\date{today}

\maketitle

\begin{abstract}
SrRuO$_3$ is a metallic ferromagnet.  Its electrical resistivity
is reported for temperatures up to 1000K; its Hall coefficient for
temperatures up to 300K; its specific heat for temperatures up
to 230K.  The energy bands have been calculated by self-consistent
spin-density functional theory, which finds a ferromagnetic ordered
moment of 1.45$\mu_{{\rm B}}$ per Ru atom.  The measured linear specific
heat coefficient $\gamma$ is 30mJ/mole, which exceeds the theoretical
value by a factor of 3.7.
A transport mean free path at room temperature
of $\approx 10 \AA$ is found.  The resistivity increases nearly linearly
with temperature to 1000K in spite of such a short mean free path that
resistivity saturation would be expected.  The Hall coefficient is small
and positive above the Curie temperature, and exhibits both a low-field
and a high-field anomalous behavior below the Curie temperature.
\end{abstract}

%\narrowtext

\section{Introduction}

SrRuO$_3$ has a nearly cubic perovskite structure \cite{Callaghan,Jones}.  The
combination of good chemical stability, good metallic conductivity \cite{Noro}, and
easy epitaxial growth on various perovskite substrates \cite{Eom2,Eom,Wu}
makes it attractive for possible multilayer device applications \cite{Antognazza}.
There is some evidence \cite{Catchen} that above $\approx$800K the
structure reverts to cubic.  Nominally single crystals are therefore 
likely to be randomly twinned arrangements of the three orientations of
orthorhombic ordering.
Below T$_c$=160K it is ferromagnetic with a moment 1.1$\mu_B$ at T=0K \cite{Kanb1},
the largest ordered moment known to arise from 4d electrons.  Assuming
a nominal valence Ru$^{4+}$ and configuration 4d(t$_{{\rm 2g}}$)$^4$,
full Hund's rule spin alignment ($\uparrow$t$_{{\rm 2g}}^3$,
$\downarrow$t$_{{\rm 2g}}^1$) predicts a moment 2$\mu_B$ and insulating
(``half-metallic'') behavior of the majority spins.  The deficiency of moment has
received three explanations, (1) collective (rather than atomic) band magnetism
\cite{Longo} (2) spin canting \cite{Gibb}, and (3) 
incomplete alignment of magnetization in different domains due
to large magnetocrystalline anisotropy and random crystallographic
domain alignment \cite{Kanb3}.  In this paper we offer new transport and 
thermodynamic data, and a band structure calculation which agrees 
with the band magnetism view of the reduced moment.  In particular, the
specific heat shows a very large Fermi level density of states, which band theory
assigns to both majority and minority spins, with majority spins slightly more
numerous.  The majority spin electrons have a small Fermi velocity, suggesting
that most of the transport currents are carried by minority spin electrons.

\section{Electronic Structure}

The electronic structure calculations were done using the self-consistent
linear muffin-tin orbital (LMTO) method \cite{lmto} in the local
spin density approximation (LSDA). The calculations for the distorted
unit cell, containing 20 atoms, were converged using 64 k-points in the
irreducible (1/8) Brillouin Zone. The density of states (DoS) and Fermi
velocities were calculated by tetrahedron integration using a finer
k-point mesh. The structural parameters were taken from ref. \cite{Shikano}.

A first calculation was done for the undistorted cubic structure
with 5 atoms per cell.
The shape of the DoS is similar to that for the real unit cell,
except for details. The DoS value per atom at the Fermi 
energy ($E_F$), is somewhat smaller than for the large cell, making the Stoner 
factor (1.04) and magnetic moment (0.6 $\mu_{{\rm B}}$) smaller. 
The total energy is lower by about 25 mRy per formula
unit (fu) in the ferromagnetic configuration compared to the paramagnetic case.
For the remaining discussion we give only the results from
the calculations for the large distorted cell.  

Fig. 1 shows the DoS for the two spins. The bands extending from about
-8eV to about -3eV relative to $E_F$ are primarily of oxygen p parentage.
There is a gap of about 1eV separating these from bands of primarily
(74\%) Ru d parentage.  The Fermi level lies about 2eV above the bottom
of these bands, filling to about two-thirds occupancy a region of high
density of states, presumably associated with the t$_{{\rm 2g}}$
orbitals which are mixed mostly through $\pi$-overlap with oxygen p states. 
The positions of these levels accord well with the assignments made by
Cox {\sl et al.} \cite{Cox} based on UPS measurements.
Above the Fermi level comes an increasing amount of Ru s and Sr s and d
levels in addition to the Ru d states of e$_{{\rm g}}$-character.
In the paramagnetic case
when the exchange splitting is zero, $E_F$ falls near the peak in the
DoS so that the Stoner factor becomes quite large, 1.39. Thus a 
ferromagnetic transition is expected, mainly due to the Ru-d band, and the
magnetic moment is 1.45 $\mu_B$ per Ru atom in the spin polarized calculation.
The other atoms have only small moments, about 0.06 and 0.14 $\mu_{{\rm B}}$ 
per Sr and O atom respectively. The total energy difference between the ferro- and
paramagnetic configuration is about 24 mRy per fu.  Our calculated moment
exceeds the zero-field measured value $\approx 1.1 \mu_{{\rm B}}$ \cite{Kanb1}.
There is some uncertainty in the true value of the saturation magnetization.
Kanbayasi \cite{Kanb3} extrapolates a high-field value as great as 2$\mu_{{\rm B}}$.
It is possible that a large magnetocrystalline anisotropy 
\cite{Kanb1,Kanb3,Klein} pins the magnetization
perpendicular to the orthorhomic c-axis of the distorted structure, which is
randomly aligned in the nominally single-crystal sample.

It is interesting that our calculated moment is so sensitive to the small
distortion of crystal structure from cubic.  This accords with the discovery
by Kirillov {\sl et al.} \cite{Kirillov} that optic phonon modes are
affected by the magnetic order.

Table I shows the calculated value of the density of states at the
Fermi surface for each atom species and spin type, and
also the calculated values of the Fermi velocity and Fermi surface
area.  
Three bands of majority spins cross the Fermi level, with
quite small Fermi velocities, whereas 8 bands of minority spin cross the
Fermi level with more ordinary size Fermi velocities.
The DoS at $E_F$, $N(E_F)$, is sensitive to changes in the magnetic moment,
because of the position of $E_F$ on the downslope of majority and upslope
on the minority spins as shown in fig. 1. At convergence we find the total 
$N(E_F)$= 190 states/(Ry$\cdot$cell), where 60 \% is majority spin.  This DoS
is concentrated on the Ru sites, making the value very high per
Ru atom.  The DoS is even higher in the paramagnetic state,
where the Fermi surface lies in a region of very small Fermi velocity.  
Ferromagnetic ordering diminishes the area of the Fermi surface
to 75\% of the paramagnetic state value, mostly by diminishing the area
of the up spin surface.  The DoS is reduced to 58\% of the paramagnetic
value.  The up spin contribution is diminished because of the reduced
area of Fermi surface, and the down spin contribution is reduced because
of the increased Fermi velocity.

	Recently we learned of an unpublished LAPW calculation
by Singh \cite{Singh}.  Near the Fermi level, Singh's results
agree very closely with ours, including the size of the ordered
moment.  Farther away there are interesting differences of detail,
but reasonable qualitative accord.

\section{Specific Heat}

Polycrystalline samples of mass $\approx$ 0.5g were prepared for 
specific heat measurements.
Stoichiometric mixtures of SrO$_2$ and RuO$_2$ (99.9\%  Atomergic Chemetals) 
were first heated for 24 h at 1000$^{\circ}$C
and refired again for 80 h at 1100$^{\circ}$C.
Several grindings were done in an agate mortar between the thermal
treatments in order to get homogeneous samples.
After 50 hours at 1150$^{\circ}$C the reactions were
fully realized, and slow quenching (10$^{\circ}$C/h)
was done under oxygen.  X-ray powder diffraction of this product
showed a single phase consistent with the previous report
of Jones {\sl et al.} \cite{Jones}.

The low temperature heat capacity measurement used a thermal
relaxation technique with 1\% accuracy \cite{Sanchez}.
The polycrystalline sample, 47mg in mass, was measured in zero
external field between 1.2 and 32K.  
Fig. 2 shows the data below 7K in the usual $C/T$ {\sl vs.} $T^2$ plot.
Below 10K the data are fit by the low $T$ approximation
\begin{equation}
\frac{C}{T}=\gamma+\frac{12}{5}R\pi^4 \frac{T^2}{\Theta_{{\rm D}}^3}
\end{equation}
where $\gamma$ is the Sommerfeld constant and $\Theta_{{\rm D}}$ the
Debye temperature in the limit $T\rightarrow 0$.
From a least squares fit between 1 and 7K we extract
the values $\gamma$=6.0mJ/atg (1 atg is 47.3g or 1/5 mole SrRuO$_3$) and
$\Theta_{{\rm D}}$=368K.  Deviations occur at higher temperature, 
indicating that $\Theta_{{\rm D}}$ has a minimum near 25K and
then increases with $T$.  Such a variation is quite common and reflects
additional modes in the 10meV range beyond those in the Debye distribution.
The measured $\gamma$ corresponds to a Fermi
Level DoS of 173 states/(Ry formula unit) (both spins) which exceeds
the band theoretical value by a factor 
$1+\lambda_{\gamma}\equiv \gamma_{exp}/\gamma_{th} = 3.7$.  
Almost an identical enhancement
of experiment over theory was recently found for the layered metal
Sr$_2$RuO$_4$ \cite{Maeno,Oguchi}.  By contrast, the rutile structure metal
RuO$_2$ has a theoretical density of states at the Fermi surface which is
only half as large per ruthenium atom \cite{Glassford}. 
Comparing with the measured specific heat \cite{Passenheim}
this corresponds to a value $\lambda_{\gamma}=0.45$ 
close to what is expected
from normal electron-phonon mass enhancement.
Thus RuO$_2$ appears to be a conventional metal, while both SrRuO$_3$
and Sr$_2$RuO$_4$ have large mass enhancements 
$\lambda_{\gamma}$ presumably caused less by electron-phonon effects than by
spin fluctuations or some other kind of Coulomb correlation effect.

Fig. 3 shows the high temperature specific heat of a 0.36g sample
cut from the same ceramic mass, and measured
in an adiabatic, continuous heating calorimeter \cite{Junod}.
The accuracy is 1\% and the scatter 0.02\%.
The full width of the ferromagnetic transition exceeds
the temperature steps used in data reduction by about two orders of
magnitude.

The total specific heat approaches 20J/(atg K) at T=250K, a
relatively low value compared to the Dulong-Petit saturation
at 24.94 J/(atg K).  Neglecting electronic contributions, this
sets a lower limit to the Debye temperature,
$\Theta_{{\rm D}}$(250K)$>$ 540K, significantly higher than the low temperature
value, 368K.  This comparison suggests that the phonon density
of states is widely dispersed, including high frequency optical
modes from ruthenium-oxygen bond stretching vibrations.

The ferromagnetic transition centered at T$_{{\rm c}} \approx$ 160K
appears as a surprisingly small step rather than the 
quasi-logarithmic divergence frequently seen.  A closer look at
the derivative of the specific heat in fig. 3 shows two transition
components at 158 and 162K, perhaps caused by sample inhomogeneity.
No significant anomaly is seen in the range 35-45K where
the resistivity shows some structure (see below.)

The entropy associated with the transition can be estimated 
by subtracting phonon and
spinless electronic contributions to the specific heat.
Most of the magnetic entropy in model systems with well-localized
spins is released in a specific heat peak extending from
$\approx$0.5T$_{{\rm c}}$ to  $\approx$1.5T$_{{\rm c}}$,
{\sl i.e.} from 80 to 240K, and should  equal $R\ln3$ or $R\ln2$
for full disordering of localized spin of 1 or 1/2 ions which are
fully ordered at $T=0$.  Since we have no clear way to make
this subtraction, we used two different methods modeled after
a critical transition and a mean-field transition respectively.
Phonons were modeled by sums of Einstein functions and electrons
were assumed to give a contribution $\gamma T$.  The results for the
integrated entropy $\Delta S$ in the interval 40 to 200K were
$\approx 1.1$ J/(atg K) and 2.7 J/(atg K) respectively by the two
methods.  These correspond to 12\% and 30\% of $R\ln3$.  Apparently
no more than 1/3 of the entropy of $S=1$ Ru spins is removed by
ferromagnetic ordering.  However, if we assume an electronic
term $\gamma T$ with $\gamma$ independent of temperature and
having the value measured at low $T$, as was done in both entropy
estimates, then there is a lot of electronic entropy ($\approx
.7R$ per Ru atom) removed by cooling from 200K to $T=0$.
It is probably not correct to attempt a complete separation between
electronic and magnetic entropy.

\section{Transport Properties}

Single crystal samples of dimensions $\approx 1 \times 1 \times 0.5$mm
were grown for resistivity measurements by slow cooling in SrCl$_2$
flux \cite{Bouchard}.
First polycrystalline SrRuO$_3$ was prepared from stoichiometric quantities
of RuO$_2$ (99.9\%) and SrCO$_3$ (99.99\%).  The single-phase product was 
mixed with SrCl$_2$ anhydrous in the weight ratio 1:30, and placed in a
50cm$^3$ platinum crucible covered with a platinum lid in a vertical furnace
with programmable temperature control.  The refractory crucible supports 
induced a small temperature gradient towards the base of the crucible.
The flux mixture was heated to 1260$^{\circ}$C (200$^{\circ}$C/h)
for homogenization of the melt, soaked for 40h, and cooled to 800$^{\circ}$C
(at 2$^{\circ}$C/h).  The crucibles were then removed and cooled quickly
to room temperature.  Crystals a few mm in size were easily separated
from the frozen flux by immersing the crucible in hot water.

Resistivity was measured in a four probe configuration on samples
of typical dimensions $\approx 1 \times 0.5 \times 0.05$mm$^3$.
Approximately 10 samples were measured.  The major features were
reproducible, {\sl i. e.} the break in slope at T$_{{\rm c}}$, the qualitative
shape, and the absolute magnitude (to within a factor of 2).
We attempted to measure the anisotropy of resistivity between
the long axis and the thin axis of the samples; the effect was less
than 10\%.  Given that LSDA band theory (section II) finds an
orthorhombic anisotropy varying from 10\% to 40\%, the lack
of measurable anisotropy of our samples may mean that there is a
random distribution of orthorhombic domains or twins in our samples.
Results for one of our samples up to 1000K are shown in fig. 4. 

The electrical resistivity has been reported previously at temperatures
below 300K.  Polycrystalline samples were reported in refs. 
\cite{Noro,Shikano,Neumeier}, single crystals in ref. \cite{Bouchard}, and oriented
thin film samples in refs. \cite{Eom,Wu,Antognazza}.  
Our data agree with all measurements
concerning the shape of $\rho(T)$ and the break in slope at T$_{{\rm c}} 
\approx$ 160K.  Measurements on polycrystalline samples tend to yield
absolute magnitudes of $\rho(T)$ about 5 times larger than seen
on single crystal or oriented film samples.  Our measurements, as
well as previous ones on single crystals or oriented films 
show a surprising sample-dependence in the low temperature
power law, with the exponent $p$ (defined by $\rho(T)-\rho(0) \propto T^p$)
appearing to vary between 1 and 2.  

Fig. 5 shows the resistivity on a different sample in the range
0 to 300K, along with the derivative $d\rho/dT$ and a Bloch-Gr\"uneisen
curve.  As is typical in metallic ferromagnets, there is a peak
in $d\rho/dT$ at the Curie point, similar
to the peak in $C(T)$.   We presume that the resistivity
can be composed of a sum of electron-phonon scattering and spin-fluctuation
scattering.  Above the Curie temperature, the spin-fluctuation part
should slowly saturate at an approximately temperature-independent value.
The Bloch-Gr\"uneisen curve in fig. 5 is chosen to match the slope
of the data at high temperature.  Insofar as this curve represents a fair
estimate of the electron-phonon contribution to $\rho(T)$, we can use
it \cite{Allen}
to extract a value of the electron-phonon coupling constant $\lambda_{{\rm tr}}$.
Lacking more detailed information, we assume that the scattering
rate is the same for up and down spin electrons, and roughly constant for
all electrons at the Fermi surface.  Then $\sigma=\Omega_p^2 \tau/4\pi$
where the Drude plasma frequency $\Omega_p^2$ is defined by
\begin{equation}
\Omega_{xx}^2=\frac{4\pi e^2}{\Omega_{{\rm cell}}}
\sum_{k\sigma} v_{k\sigma x}^2 \delta(\epsilon_{k\sigma}-\epsilon_F).
\end{equation}
From the values given in table I, the $xx$, $yy$, and $zz$ components
of the Drude plasma frequency tensor have the values 3.0, 2.5, and 2.3eV
in the ferromagnetic state, and 2.2, 2.5, and 2.3eV in the paramagnetic
state.
For a random mixture of orthorhombic domains we should use the root mean
square value, $\Omega_p$=2.6eV (ferromagnetic) or
2.3eV (paramagnetic).  Using a Debye spectrum with $\Theta_D$
=368K, the curve of fig. 5 corresponds to $\lambda_{{\rm tr}}$=0.5,
calculated using the paramagnetic state value of the Drude plasma
frequency.  This is similar to the value found in ref. \cite{Glassford}
for RuO$_2$.  However, because LSDA band theory agrees
well with the specific heat $\gamma$ in the latter compound but not
in SrRuO$_3$, the value of $\lambda_{{\rm tr}}$ found here 
cannot be regarded as particularly reliable. 

It is surprising how rapidly
thermal scattering sets in as temperature increases from $T$=0,
much more rapidly than the Bloch $T^5$-law, but sample-dependent
as mentioned above.
We do not understand the origin of either the rapid rise of $\rho(T)$
or its sample dependence.  Independent of a model for the scattering,
however, we can estimate the carrier mean free path $\ell$, using the equation
\cite{Allen}
\begin{equation}
\sigma=\frac{e^2}{24\pi^3 \hbar}[A_{FS\uparrow}\ell_{\uparrow}
+A_{FS\downarrow}\ell{\downarrow}]
\end{equation}
where $A_{FS}$ is the Fermi surface area.
This result depends only on the assumption of not having too strong
a variation of $\ell_k$ from point to point on the Fermi surface.
For paramagnetic materials with $A_{FS\uparrow}=A_{FS\downarrow}$,
eqn. 3 permits a robust estimate of the average mean free path, since
LDA band theory generally gets the shape of the Fermi surface to
very good accuracy,  
not contaminated by renormalization
or other likely errors of LDA.  
In the ferromagnetic case, we are less certain about the
correctness of LSD theory concerning spin-splitting;
therefore the separate
values of $A_{FS\uparrow}$ and $A_{FS\downarrow}$
are less robust.
Using the room temperature resistivity of 200$\mu\Omega$cm for the
sample shown in fig. 5, and the paramagnetic value of the Fermi
surface area, the room temperature mean free path is 7$\AA$,
as small as is found in good superconductors with the A15 structure.
However, unlike the case of A15 superconductors, there is no
sign of ``saturation'' of the resistivity ($\rho(T)$ approaching
a temperature-independent value.)  If the value $\rho$(300K)
=150$\mu\Omega$cm from fig. 4 is used, $\ell$ becomes $\approx 10\AA$.
This is still short enough that the Boltzmann quasiparticle gas theory
should not be applicable \cite{Allen}.

The origin of this short room temperature mean free path
is the small Fermi velocity found by LDA band theory in 
the paramagnetic phase.
One way of having a larger mean free path is to argue that
the paramagnetic state has fluctuating spins with medium range
order, so that locally for time intervals not too long the
energy bands are really the ferromagnetic bands.
The areas of the up and
down spin Fermi surfaces are not very different.  However, assuming
a roughly constant time between scattering events, the small velocity
of the majority spin carriers suggests a significantly smaller mean
free path for the up spins than for the down spins.  Therefore,
we assume that the second term of eq. 3 dominates, and find from
the data of fig. 5 an upper limit $\ell_{\downarrow}$(300K)=17$\AA$,
or 22$\AA$ using the data of fig. 4.
The up spin electrons are unlikely to have a mean free path larger
than $\approx 9\AA$. 
By $T=500$K, even the down spin electrons
should have $\ell < 10\AA$, and Boltzmann theory should 
fail visibly.  However, the resistivity as shown in fig. 4 
continues to rise almost linearly for $T$ up to 1000K.  The origin
of this behavior is not known, but is reminiscent of the resistivity
seen in high T$_{{\rm c}}$ superconductors and seen also in the
high $T$ metallic phase of VO$_2$ \cite{Allen2}.

Our samples show a small but significant enhancement in $d\rho/dT$
in the range 35-45K (fig. 5).  This effect is sample-dependent.
In exactly this range, Kanbayasi \cite{Kanb2}
found a dramatic change in magnetocrystalline anisotropy and presumed
change in orientation of the magnetization in certain samples 
called ``tetragonal'' rather than ``pseudo-cubic'' (orthorhombic).
However, no sign of anomaly can be seen in the specific heat measured
on polycrystalline samples in this range (fig. 3).  Possibly our
nominally single crystal samples actually contain
varying amounts of Kanbayasi's second crystal structure.

The Hall effect was measured in a six-probe configuration,
with a typical current of 10mA and magnetic field up to 12T.
The sign of the Hall voltage was calibrated with a sample of
Bi$_2$Sr$_2$CaCu$_2$O$_8$ superconductor.  
Fig. 6 shows typical curves of Hall voltage versus field at
temperatures above and below the Curie temperature.
Above T$_{{\rm c}}$ at 190K there is a normal weak positive
Hall coefficient, $\approx 2 \times 10^{-10}$m$^3$/C, which diminishes
gradually at higher $T$.  Just below T$_{{\rm c}}$
at 140K, the data could be interpreted as showing the sum of
an anomalous Hall effect of positive sign and an ordinary Hall
effect of negative sign.  However, the crossover field 
($\approx$ 4T) is much too
large.  The data at 100K show clearly that there are three regimes
of field, with an anomalous Hall effect occuring in very weak fields,
not resolved in our experiment, and then two higher field regimes,
each characterized by approximately linear behavior of $V_H$ with
applied field, but with a slope change near 6T.

Since the Fermi surface contains both hole-like and
electron-like regions, it is very difficult to predict from the
band structure how $R_H$ should behave.  The temperature variation
can be rationalized in a model where electron-phonon scattering
is dominant on the hole-like parts and spin-fluctuation scattering
dominates on electron-like parts.  As temperature increases for
T$>$T$_{{\rm c}}$, electron-phonon scattering increases while spin
fluctuation scattering saturates.  This drives conductivity
increasingly onto the electron-like parts of Fermi surface.

\acknowledgements

We thank C. H. Ahn, P. Almeras, L. Mi\'eville, J.-M. Triscone,
Z.-G. Ye, and K. Yvon for helpful discussions.
PBA was supported in part by U. S. NSF grant no. DMR-9417755.

\begin{figure}
\caption{Electronic density-of-states of ferromagnetic SrRuO$_3$.  Majority
spin is plotted upward, minority spin downward.  The cell contains four
formula units.}
\end{figure}

\begin{figure}
\caption{Low temperature specific heat C/T plotted versus T$^2$ for SrRuO$_3$}
\end{figure}

\begin{figure}
\caption{Specific heat $C(T)$ (diamonds) and derivative $dC(T)/dT$ 
(full line; note the scale factor) for SrRuO$_3$ in the 15-250K range}
\end{figure}

\begin{figure}
\caption{Resistivity versus temperature for SrRuO$_3$}
\end{figure}

\begin{figure}
\caption{Resistivity versus temperature for SrRuO$_3$ from 0 to 300K, with
a Bloch-Gr\"uneisen curve (solid line) for comparison, and $d\rho/dT$ in
arbitrary units}
\end{figure}

\begin{figure}
\caption{Hall voltage versus field for SrRuO$_3$ at three temperatures}
\end{figure}

\begin{table}
\begin{tabular}{|lc|cccc|ccc|c|} 
 & & \multicolumn{4}{c|}{DoS [states/(Ry cell)]} &
        $\langle v^2_{F,x} \rangle$ & $\langle v^2_{F,y} \rangle$ & 
        $\langle v^2_{F,z} \rangle$ & Area of FS\\
 & & Sr & Ru & $\rm O_3$ & Total & \multicolumn{3}{c|}
        {[$10^{14}$ $\rm cm^2/s^2$]} & [$10^{17}$ $\rm cm^{-2}]$\\ \hline
\multicolumn{2}{|c|}{Paramagnetic} & 12. & 246. & 67. & 325. & 0.6 & 0.8 & 0.7
        & 1.06 \\ \hline
 & Up & 5. & 84. & 25. & 114. & 1.0 & 0.6 & 0.6 & 0.68 \\
Magnetic & Down & 4. & 55. & 17. & 76. & 3.5 & 2.6 & 2.0 & 0.91 \\ \cline{2-10}
 & Total & 9. & 139. & 42. & 190. & 2.0 & 1.4 & 1.2 & 1.59 \\ 
\end{tabular}
\caption{Calculated properties of $\rm SrRuO_3$}
\end{table}

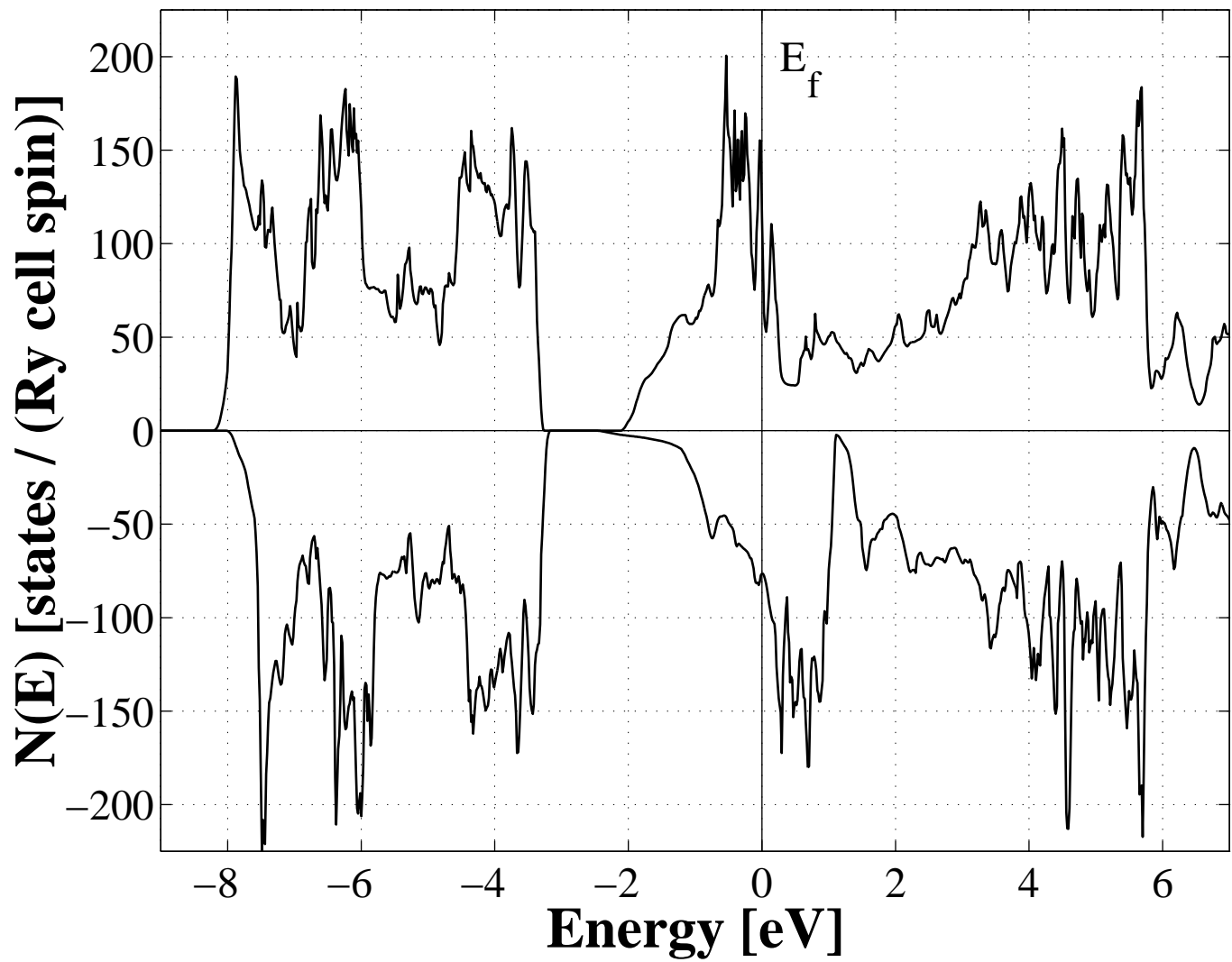
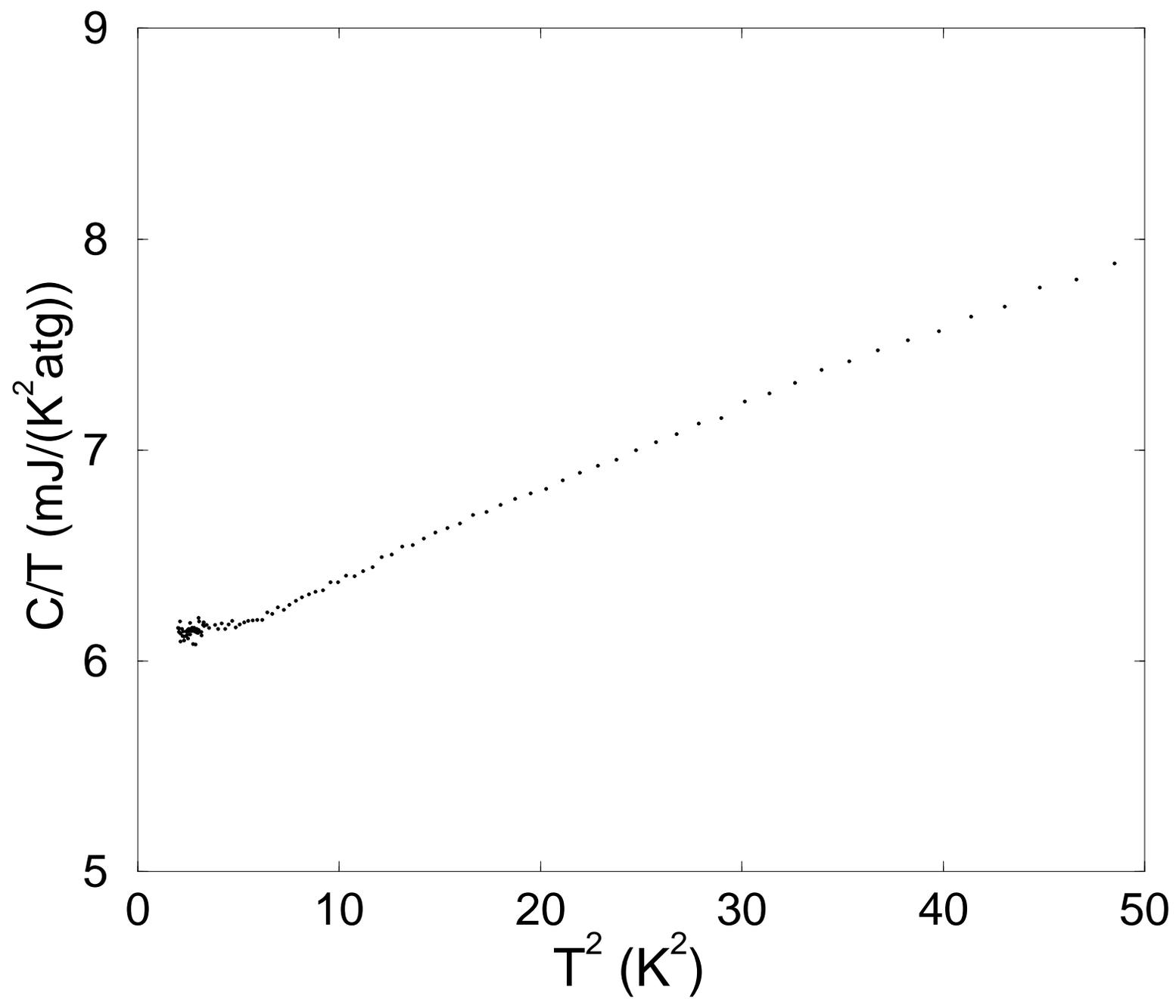
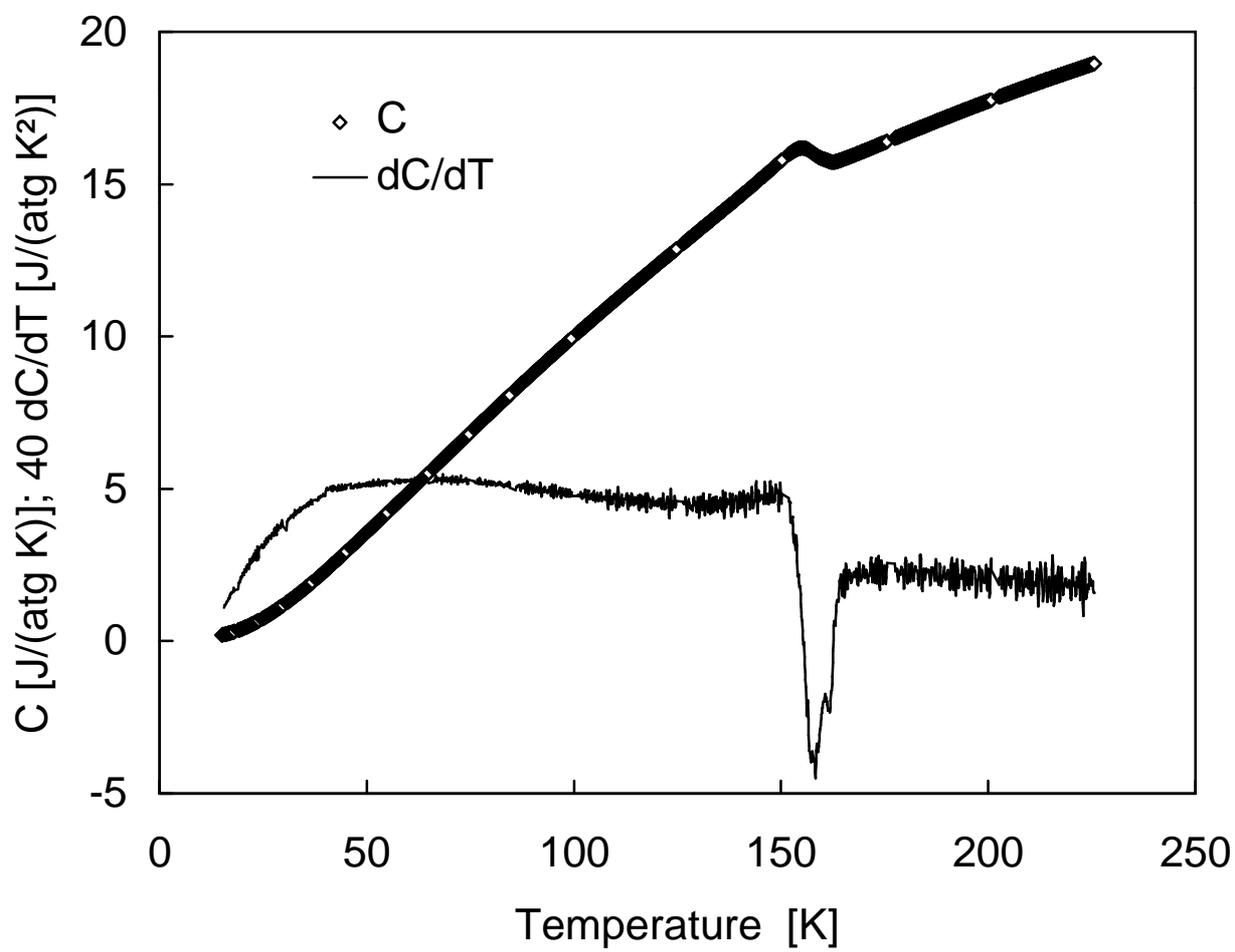
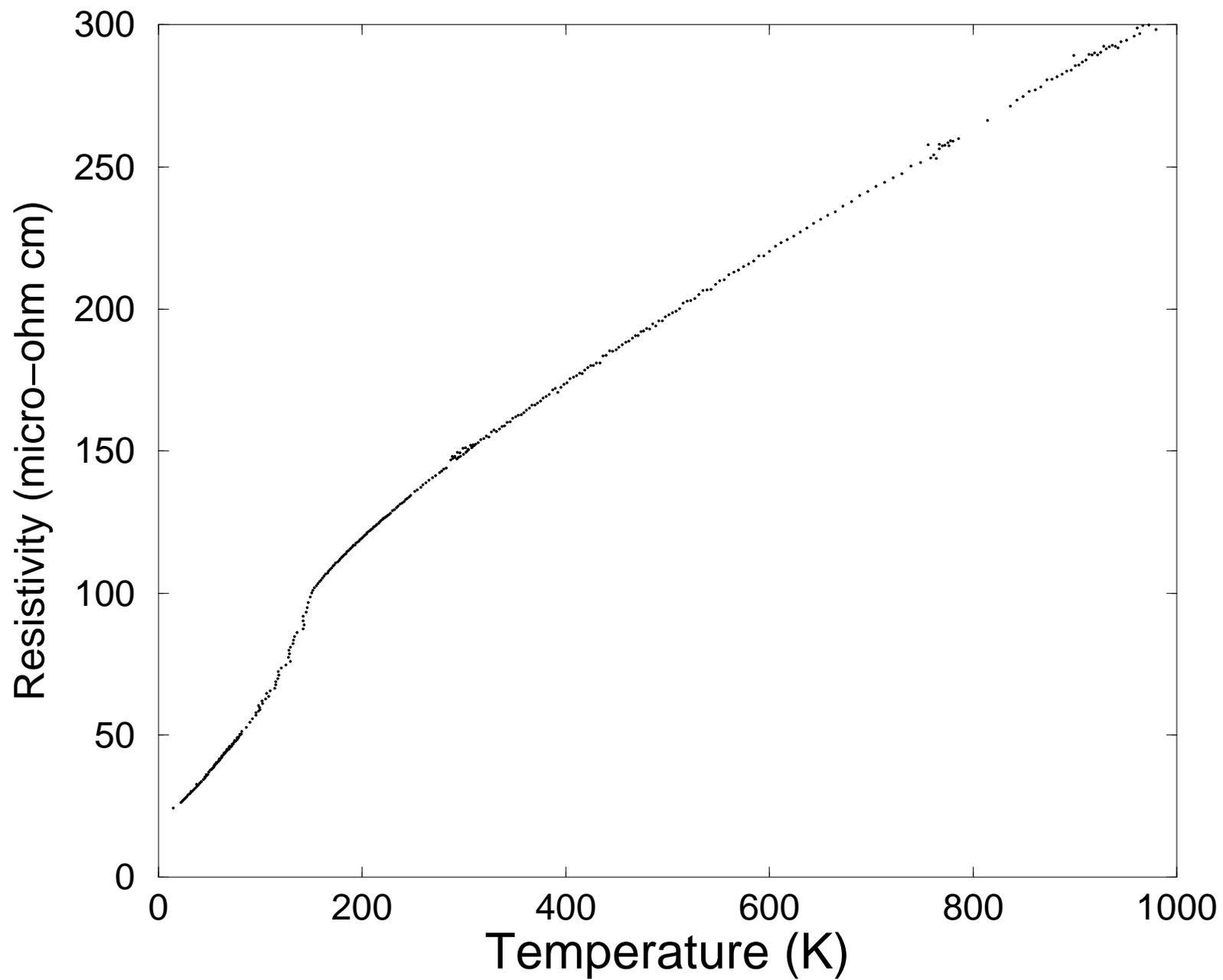
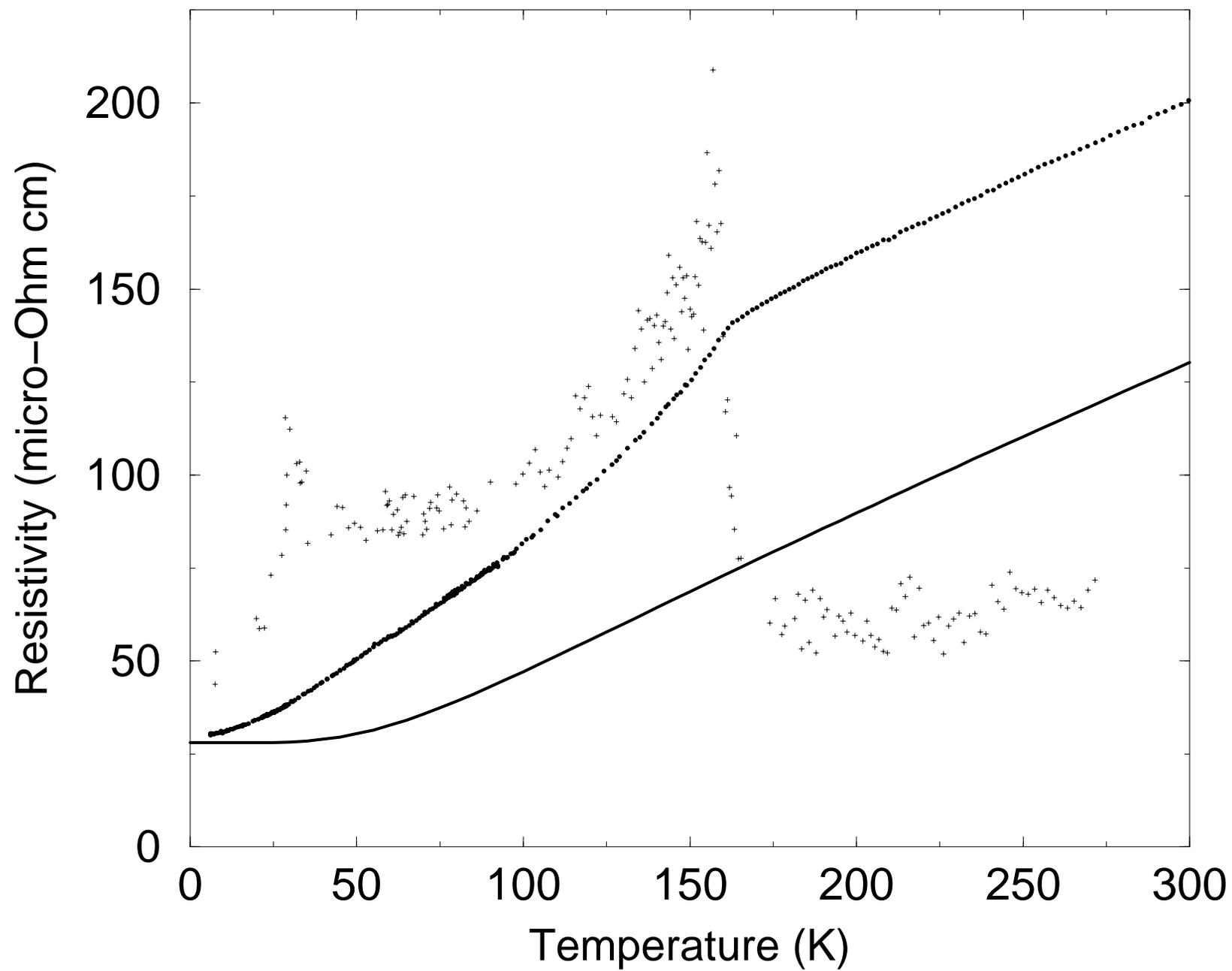
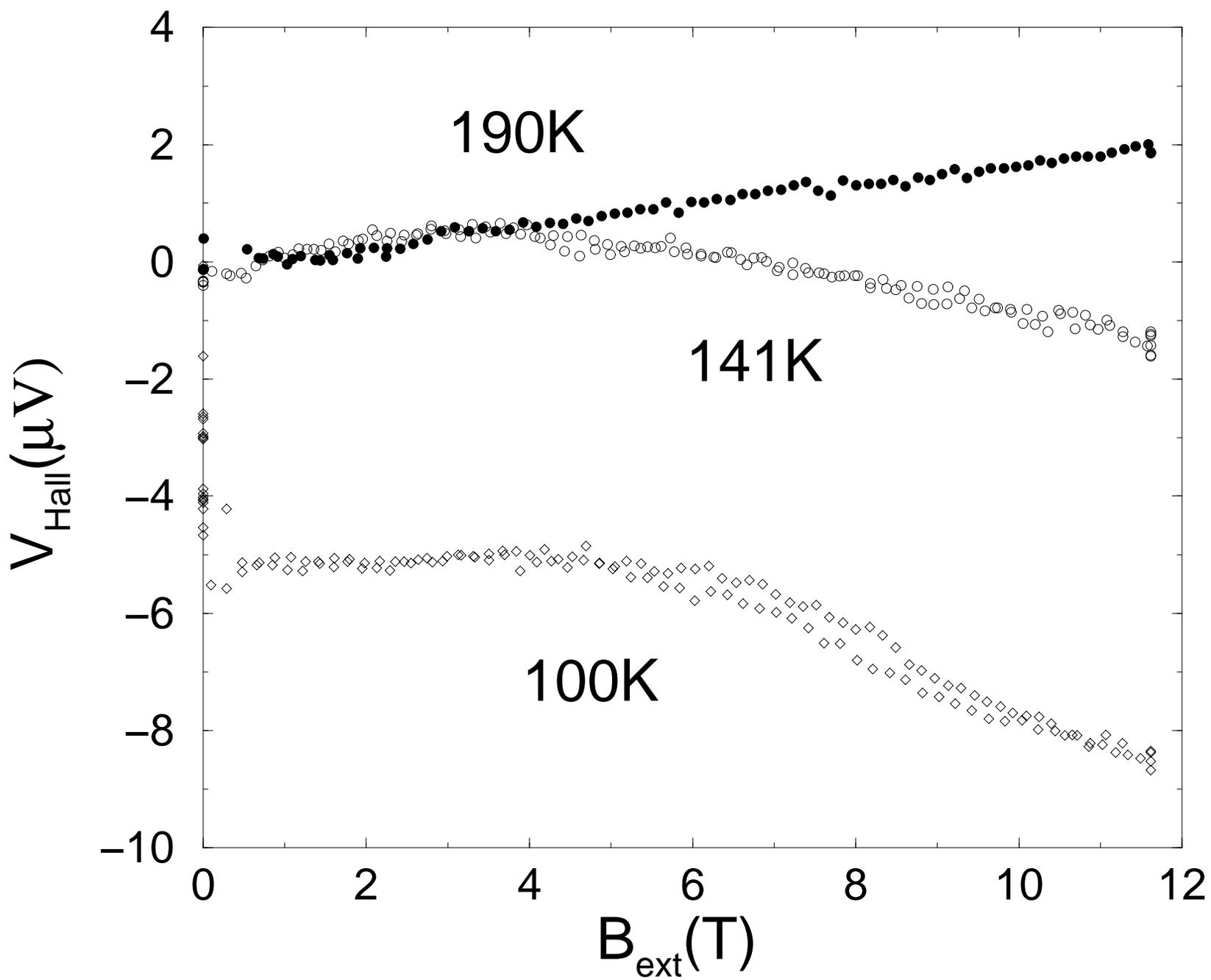
\end{document}